\documentclass[a4paper,12pt]{article}

\usepackage{indentfirst}

\usepackage{amsfonts,amsmath,amssymb,bm,a4wide,graphicx,cite,makeidx,multicol}

\begin{document}

\title{Spin Light of Neutrino in Dense Matter}

\author{A.Grigoriev\footnote{ax.grigoriev@mail.ru}, \
        A.Studenikin \footnote{studenik@srd.sinp.msu.ru },
   \\
   \small {\it Department of Theoretical Physics,}
   \\
   \small {\it Moscow State University,}
   \\
   \small {\it 119992 Moscow,  Russia }
   \\
   A.Ternov \footnote {a\_ternov@mail.ru}
   \\
   \small{\it Department of Theoretical Physics,}
   \\
   \small {\it Moscow Institute for Physics and Technology,}
   \\
   \small {\it 141700 Dolgoprudny, Russia }}

\date{}
\maketitle

\begin{abstract}
    We develop the quantum theory of the spin light of neutrino ($SL\nu$)
    exactly accounting for the effect of background matter. Contrary
    to the already performed studies
    of the $SL\nu$, in this paper we derive explicit and closed expressions for the
    $SL\nu$ rate and power and for the emitted photon energy,
    which are valid for an arbitrary matter density
    (including very high values). The
    spatial distribution of the radiation power and the dependence
    of the emitted photon energy on the direction of radiation
    are also studied in detail for the first time. We analyze the
    $SL\nu$ polarization properties and show
    that within a wide range of neutrino momenta and matter densities
    the $SL\nu$ radiation is circularly
    polarized. Conditions for effective $SL\nu$ photon
    propagation in the electron plasma are discussed. It is also
    shown
    that in  dense matter the average energy of the emitted
    photon can reach  values in the range from one third of the
    neutrino momentum up to one half of the neutrino energy in matter.
    The main features of the studied radiation are summarized,
    and possibilities for the $SL\nu$ production during different
    astrophysical and cosmology processes are discussed.

\end{abstract}

\section{Spin light of a neutrino in matter}

There exist various mechanisms for the production of
electromagnetic radiation by a massive neutrino moving in a
background environment (see, for instance,
\cite{IoanRafPRD97})\footnote {A brief classification of the known
mechanisms of the electromagnetic radiation by a neutrino is given
in the first paper of \cite{LobStuPLB03_04}.}. We have recently
shown \cite{LobStuPLB03_04} within the quasi-classical approach,
that  a massive neutrino moving in background matter can emit a
new type of electromagnetic radiation. This radiation has been
termed  the "spin light of neutrino" ($SL\nu$) in matter. In
\cite{DvoGriStuIJMPD05} we have also considered $SL\nu$ in
gravitational fields of rotating astrophysical objects. Developing
the quantum theory of this phenomenon
\cite{StuTerPLB05,StuTerQUARKS_04_0410296}, we have demonstrated
that $SL\nu$ arises owing to two underlying phenomena: (i) the
shift of neutrino energy levels in matter, that are different for
the two opposite neutrino helicity states, and (ii) the emission
of an $SL\nu$ photon in the process of neutrino transition from
the "excited" helicity state to the low-lying helicity state in
matter. However, calculations of the transition rate and radiation
power have been performed in the limit of a low matter density
and, therefore, evaluation of a consistent quantum theory of
$SL\nu$ still remains an open issue.

In this paper we develop the quantum theory of $SL\nu$, exactly
taking into account the effect of background matter, and obtain
expressions for the $SL\nu$ rate and power that are valid for any
value of the matter density parameter (see also
\cite{GriStuTerCOSMION04_hep_ph0502231}). In Section 2 we briefly
discuss the modified Dirac equation and the neutrino energy
spectrum in the presence of matter which are then used (Section 3)
for derivation of the $SL\nu$ transition rate and power. We get an
exact expression for the emitted photon energy as a function of
the initial neutrino energy and the matter density parameter. The
dependence of the photon energy  on the direction of the photon
propagation is  analyzed, and a detailed study of the radiation
spatial distribution is also performed. We also derive the exact
and closed expressions for the rate and total radiation power of
$SL\nu$ and analyze them for different limiting cases. The $SL\nu$
polarization properties are studied in Section 4, and the
conclusion is made concerning the total circular polarization of
the emitted photons. Section 5 is devoted to the discussion of
restrictions on the propagation of $SL\nu$ photons that can be set
by the electron plasma. In conclusion (Section 6) we give a
summary of the investigated properties of $SL\nu$ in matter. The
$SL\nu$ production during processes of collapse and coalescence of
neutron stars or a neutron star being "eaten up" by the black hole
at the center of our Galaxy are also discussed as one of possible
mechanisms of gamma-rays production.

\section{The modified Dirac equation in matter}

To account for the influence of background matter on neutrinos we
use the approach \cite{StuTerPLB05} (similar to the Furry
representation in quantum electrodynamics) that is based on the
exact solutions of the modified Dirac equation for a neutrino in
matter:
\begin{equation}\label{Dirac} \Big\{
i\gamma_{\mu}\partial^{\mu}-\frac{1}{2}
\gamma_{\mu}(1+\gamma_{5})f^{\mu}-m \Big\}\Psi(x)=0.
\end{equation}
In the case of matter composed of electrons
\begin{equation}\label{f}
  f^\mu={G_F \over \sqrt2}(1+4\sin^2 \theta
_W) j^\mu,
\end{equation}
where the electron current $j^{\mu}$ is given by
\begin{equation}
j^\mu=(n,n{\bf v}). \label{j}
\end{equation}
Here $\theta _{W}$, $n$ and ${\bf v}$ are, respectively, the
Weinberg angle, the number density of background electrons and the
speed of the reference frame in which the mean momentum of the
electrons is zero. As it has been shown \cite{StuTerPLB05} the
solutions of Eq.(\ref{Dirac}) are given by
\begin{equation}\label{wave_function}
  \Psi_{\varepsilon,{\bf p},s}({\bf r},t) =
\frac{e^{-i( E_{\varepsilon}t-{\bf p}{\bf r})}}{2L^{\frac{3}{2}}}
  \left(
   \begin{array}{cccc}
      {\sqrt{1+ \frac{m}{E_{\varepsilon}-\alpha m}}}
      \sqrt{1+s\frac{p_3}{p}}
      \\
      {s \sqrt{1+ \frac{m}{ E_{\varepsilon}-\alpha m}}}
      \sqrt{1-s\frac{p_3}{p}}\ \ e^{i\delta}
      \\
      {s\varepsilon\sqrt{1- \frac{m}{ E_{\varepsilon}-\alpha m}}}
      \sqrt{1+s\frac{p_3}{p}}
      \\
      {\varepsilon\sqrt{1- \frac{m}{ E_{\varepsilon}-\alpha m}}}
      \sqrt{1-s\frac{p_3}{p}}\ e^{i\delta}
   \end{array}
\right),
\end{equation}
where energy spectrum is
\begin{equation}\label{Energy}
  E_{\varepsilon}=\varepsilon{\sqrt{{\bf p}^{2}\Big(1-s\alpha \frac{m}{p}\Big)^{2}
  +m^2} +\alpha m},
\end{equation}
\begin{equation}\label{alpha}
  \alpha=\frac{1}{2\sqrt{2}}{\tilde G}_{F}\frac{n}{m} \ , \ \
  \tilde{G}_{F}={G}_{F}
  (1+4\sin^2 \theta_W).
\end{equation}
In equations (\ref{wave_function})-(\ref{alpha}) $m$, ${\bf p}$
and $s=\pm 1$ are the neutrino mass, momentum and helicity,
respectively. The quantity $\varepsilon=\pm 1$ splits the
solutions into two branches that in the limit of vanishing matter
density, $\alpha\rightarrow 0$, reproduce the positive and
negative-frequency solutions for the Dirac equation in vacuum.
\begin{figure}[t]
\begin{center}
\includegraphics[scale=.7]
{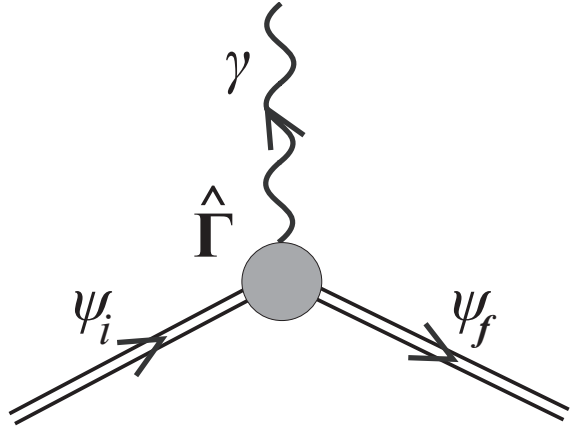} \caption{The effective diagram of the $SL\nu$ photon
emission process. The broad lines correspond to the initial and
final neutrino states in the background matter.}
\end{center}
\end{figure}

Note that generalization to the case of matter composed of
different types of fermions is  straightforward
\cite{StuTerPLB05}, and the correct value for the neutrino energy
difference corresponding to the Mikheyev-Smirnov-Wolfenstein
effect \cite{WolPRD78MikSmiYF85} can be recovered from
(\ref{Energy}). The modified effective Dirac equations for a
neutrino interacting with various background environments within
different models were previously used \cite{ChaZiaPRD88} in a
study of the neutrino dispersion relations, neutrino mass
generation and for derivation of the neutrino oscillation
probabilities in matter. On the same basis, the neutrino decay
into an antineutrino and a light scalar particle (majoron), as
well as the corresponding process of the majoron decay into two
neutrinos or antineutrinos, were studied in the presence of matter
\cite{BerVysBerSmiPLB87_89}.

\section{The $SL\nu$ transition rate and power}
The $SL\nu$ amplitude calculated within the developed quantum
theory is given by (see also \cite{StuTerPLB05})

\begin{equation}\label{amplitude}
\begin{array}{c} \displaystyle
  S_{f i}=-\mu \sqrt{4\pi}\int d^{4} x {\bar \psi}_{f}(x)
  ({\hat {\bf \Gamma}}{\bf e}^{*})\frac{e^{ikx}}{\sqrt{2\omega L^{3}}}
   \psi_{i}(x),
   \\
   \\
   \hat {\bf \Gamma}=i\omega\big\{\big[{\bf \Sigma} \times
  {\bm \varkappa}\big]+i\gamma^{5}{\bf \Sigma}\big\},
\end{array}
\end{equation}
where $\mu$ is the neutrino magnetic moment, $\psi_{i}$ and
$\psi_{f}$ are the exact solutions of  equation (\ref{Dirac}) for
the initial and final neutrino states \cite{StuTerPLB05},
$k^{\mu}=(\omega,{\bf k})$ and ${\bf e}^{*}$ are the photon
momentum and polarization vector, ${\bm \varkappa}={\bf
k}/{\omega}$  is the unit vector pointing in the direction of the
emitted photon propagation.

Integration over time in (\ref{amplitude})  yields
\begin{equation}\label{amplitude_integrated}
   S_{f i}=
  -\mu {\sqrt {\frac {2\pi}{\omega L^{3}}}}
  2\pi\delta(E'-E+\omega)
  \int d^{3} x {\bar \psi}_{f}({\bf r})({\hat {\bf \Gamma}}{\bf e}^{*})
  e^{i{\bf k}{\bf r}}
   \psi_{i}({\bf r}),
\end{equation}
where the delta-function stands for energy con\-ser\-vation, $E$
and $E'$ are the energies of the initial and final neutrino states
in matter. Performing integration over the spatial co-ordinates,
we can recover the delta-functions for the three com\-ponents of
the momentum. Finally, we get the law of energy-momentum
conservation for the considered process,
\begin{equation}\label{e_m_con}
    E=E'+\omega, \ \ \
    {\bf p}={\bf p'}+{\bm \varkappa},
    \end{equation}
where ${\bf p}$ and ${\bf p'}$ are the initial and final neutrino
momenta, respectively.
 From (\ref{e_m_con}) it follows that
the emitted photon energy $\omega$ exhibits a critical dependence
on the helicities of the initial and final neutrino states. In the
case of electron neutrino moving in  matter composed of electrons
$\alpha$ is positive. It follows that $SL\nu$ can arise only when
the neutrino initial and final states are characterized by
$s_{i}=-1$ and $s_{f}=+1$, respectively. One can also conclude
that in the process considered  the relativistic left-handed
neutrino is converted to the right-handed neutrino. A discussion
of the main properties of $SL\nu$ emitted by different flavor
neutrinos moving in matter composed of electrons, protons and
neutrons can be found in \cite{StuTerPLB05} (see also
\cite{Lob_hep_ph_0411342}).

The emitted photon energy in the considered case ($s_i=-s_f=-1$),
obtained as an exact solution of equations (\ref{e_m_con}), is
\begin{equation}\label{omega1}
\omega =\frac{2\alpha mp\left[ (E-\alpha m)-\left( p+\alpha
m\right) \cos \theta \right] }{\left( E-\alpha m-p\cos \theta
\right) ^{2}-\left( \alpha m\right) ^{2}},
\end{equation}
where $\theta$ is the angle between ${\bm \varkappa}$ and the
direction of the initial neutrino propagation. The photon energy
is a rather complicated function of the neutrino energy $E$ and
momentum $p$, the matter density parameter $\alpha$ and the angle
$\theta$.  Fig.1 shows the angular dependence of the photon energy
for different values of the neutrino momentum. From this figure
one may expect that in the case of relativistic neutrinos ($p \gg
m$) and not very dense matter ($\alpha \ll \frac {p}{m}$) $SL\nu$
is collimated along the direction of the neutrino momentum $p$
(see the dashed and solid-dashed curves). On the contrary, in the
case of non-relativistic neutrinos ($p \ll m$) and $\alpha \gg
\frac {p}{m}$ (see solid line in Fig.1) the emitted photon energy
in the direction of the neutrino momentum $p$ is suppressed. It
should also be  noted that for all cases shown in Fig.1  the
energies $\omega$ of the photons radiated at large angles $\theta$
are of the order of $\sim \alpha m$.
\begin{figure}[t]
\begin{center}
\includegraphics[scale=.5]{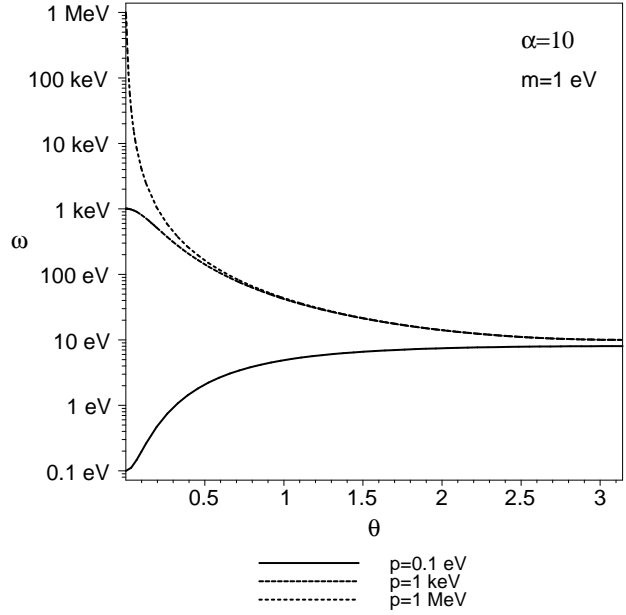}
\caption{Angular dependence of photon energy for different values
of the neutrino momentum: 1) the solid line corresponds to $p=0.1
\ eV$,  2) the dashed-solid line corresponds to $p=1 \ keV$ and 3)
the dashed line corresponds to $p=1 \ MeV$. The matter density
parameter is $\alpha =10$.}
\end{center}
\end{figure}
In the case of a not very high density of matter, when the
parameter $\alpha\ll 1$,  one can expand the photon energy
(\ref{omega1}) over $\alpha$ and in the linear approximation get
the result of \cite{StuTerPLB05,StuTerQUARKS_04_0410296}:
\begin{equation}\label{omega_2}
    \omega=
    \frac {1}{1-\beta \cos
    \theta}\omega_0,
\end{equation}
where
\begin{equation}\label{omega_0}
\omega_0= \frac {{\tilde G}_{F}} {\sqrt{2}}n \beta, \ \ \
\beta=\frac{p}{\sqrt{p^2+m^2}}.
\end{equation}

Using the expressions for the amplitude
(\ref{amplitude_integrated}) and for the photon energy
(\ref{omega1}) we calculate the spin light transition rate and
total radiation power exactly accounting for the matter density
parameter:
\begin{eqnarray}\label{Gamma}
 \Gamma &=&\mu ^2
 \int_{0}^{\pi }\frac{\omega ^{3}}{1+\tilde\beta ^{\prime
}y}S\sin \theta d\theta,
\end{eqnarray}
\begin{equation}\label{power}
I=\mu ^2\int_{0}^{\pi }\frac{\omega ^{4}}{1+\tilde\beta
^{\prime}y}S\sin \theta d\theta,
\end{equation}
where
\begin{equation}\label{S}
S=(\tilde\beta \tilde\beta ^{\prime }+1)(1-y\cos \theta
)-(\tilde\beta +\tilde\beta ^{\prime }) (\cos \theta -y).
\end{equation}
Here we introduce the notations
\begin{equation}\label{beta}
\tilde \beta =\frac{p+\alpha m}{E-\alpha m}, \ \ \tilde \beta
^{\prime }=\frac{p^{\prime }-\alpha m}{E^{\prime }-\alpha m},
\end{equation}
where the final neutrino energy and momentum are, respectively,
\begin{equation}
E^{\prime }=E-\omega , \ \ \ p^{\prime }=K\omega -p,
\end{equation}
 and
\begin{equation}
y=\frac{\omega -p\cos \theta }{p^{\prime }}, \ \ K=\frac{E-\alpha
m-p\cos \theta }{\alpha m}.
\end{equation}
Performing the integration in (\ref{Gamma}), we obtain for the
$SL\nu$ rate in matter
\begin{eqnarray}\label{gamma}
\Gamma &=&\frac{1}{2\left( E-p\right) ^{2}\left( E+p-2\alpha
m\right) ^{2}\left( E-\alpha m\right) p^{2}} \notag
\\
&&\times \left\{ \left( E^{2}-p^{2}\right) ^{2}\left(
p^{2}-6\alpha ^{2}m^{2}+6E\alpha m-3E^{2}\right) \left( \left(
E-2\alpha m\right) ^{2}-p^{2}\right) ^{2}\right. \notag
\\
&&\times \ln \left[ \frac{\left( E+p\right) \left( E-p-2\alpha
m\right) }{\left( E-p\right) \left( E+p-2\alpha m\right) }\right]
+4\alpha mp\left[ 16\alpha ^{5}m^{5}E\left( 3E^{2}-5p^{2}\right)
\right.
\\
&&-8\alpha ^{4}m^{4}\left( 15E^{4}-24E^{2}p^{2}+p^{4}\right)
+4\alpha ^{3}m^{3}E\left( 33E^{4}-58E^{2}p^{2}+17p^{4}\right)
\notag
\\
&&-2\alpha ^{2}m^{2}\left( 39E^{2}-p^{2}\right) \left(
E^{2}-p^{2}\right) ^{2}+12\alpha mE\left( 2E^{2}-p^{2}\right)
\left( E^{2}-p^{2}\right) ^{2} \notag
\\
&&-\left. \left. \left( 3E^{2}-p^{2}\right) \left(
E^{2}-p^{2}\right) ^{3} \right] \right\}, \notag
\end{eqnarray}
where the energy of the initial neutrino is given by
(\ref{Energy}) with $\varepsilon=-s_i=1$.

As it follows from (\ref{gamma}), the $SL\nu$ rate is a rather
complicated function of neutrino momentum $p$ and mass $m$, it
also non-trivially depends  on the matter density parameter
$\alpha$. In the limit of a low matter density, $\alpha\ll 1$, we
get
\begin{equation}\label{gamma_1}
\Gamma \simeq \frac{64}{3}\frac{\mu^2 \alpha ^{3}p^{3}m}{E_{0}},
\end{equation}
where $E_{0}=\sqrt{p^{2}+m^{2}}$. The obtained expression is in
agreement with our results of \cite{LobStuPLB03_04,StuTerPLB05,
StuTerQUARKS_04_0410296}. Note that the considered limit of
$\alpha\ll 1$ can be appropriate even for a very dense media of
neutron stars with $n\sim 10^{33} \ \ cm^{-3}$ because
$\frac{1}{2\sqrt{2}}{\tilde G}_{F}n\sim 1 \ eV$ for a medium
characterized by $n=10^{37} \ cm^{-3}$.

Performing also the integration in (\ref{power}), we obtain the
total $SL\nu$ radiation power in matter
\begin{align}
\label{totalpower}  I = & \frac{5}{2\left( E-p\right)^{3}\left(
E+p-2\alpha m\right)^{3}p^{2}} \times \
   \Bigg\{
      (E+p)^2(E-m)^3(E+p-2\alpha m)^3 \notag
\\    & \times (E-p-2\alpha m)^2\Big(2\alpha^2 m^2-2\alpha m(E+\frac{1}{5}p)+E^2-\frac{3}{5}p^2\Big) \notag
\\    & \times \ln
         \left(
             \frac{(2\alpha m-p-E)(E-p)}{(2\alpha m+p-E)(E+p)}
          \right)\notag
\\      - & 4\alpha mp
          \left(32\alpha^6m^6
             \Big(
                E^4-pE^3-\frac{5}{3}p^2E^2 \right.+
                \frac{5}{3}p^3E+\frac{8}{15}p^4
             \Big)\notag
\\        - & 96\alpha^5m^5
             \Big(
               E^5-\frac{23}{30}pE^4-\frac{83}{45}p^2E^3+
                \frac{11}{9}p^3E^2+\frac{38}{45}p^4E-\frac{1}{10}p^5
             \Big) \notag
\\        + & 128\alpha^4m^4
             \Big(
                E^6-\frac{47}{80}pE^5-\frac{511}{240}p^2E^4 
                +\frac{127}{120}p^3E^3+
                \frac{157}{120}p^4E^2
                -\frac{89}{240}p^5E-\frac{7}{48}p^6
             \Big)\notag
\\        - & 96(E^2-p^2)\alpha^3m^3
             \Big(
                E^5-\frac{53}{120}pE^4
                -\frac{3}{2}p^2E^3+\frac{89}{180}p^3E^2
                +\frac{47}{90}p^4E-\frac{19}{360}p^5
            \Big) \notag
\\        + & 42(E^2-p^2)^2\alpha^2m^2
             \Big(
                E^4-\frac{32}{105}pE^3
                -\frac{314}{315}p^2E^2+
                \frac{4}{21}p^3E+\frac{17}{105}p^3
             \Big) \notag
\\        - & 10\alpha m(E^2-p^2)^3
             \Big(
                E^3-\frac{4}{25}pE^2
                -\frac{17}{25}p^2E+\frac{2}{25}p^3
             \Big)
        +  \left.(E^2-p^2)^4
             \Big(
                E^2-\frac{3}{5}p^2
             \Big)
          \right)
   \Bigg\}.
\end{align}
In the case $\alpha\ll 1$, we get
\begin{equation}\label{power_1}
I\simeq \frac{128}{3}\mu ^{2}\alpha ^{4}p^{4}
\end{equation}
in agreement with Refs.\cite{LobStuPLB03_04,StuTerPLB05,
StuTerQUARKS_04_0410296}.

Let us consider the $SL\nu$ rate and power for the different
limiting values of the neutrino momentum $p$ and matter density
parameter $\alpha$. In the relativistic case $p\gg m$ from
(\ref{gamma}) we get
\begin{equation}\label{p_gg}
\Gamma = \left\{
  \begin{tabular}{c}
  \ $\frac{64}{3} \mu ^2 \alpha ^3 p^2 m,$ \\
  \ $4 \mu ^2 \alpha ^2 m^2 p$, \\
  \ $4 \mu ^2 \alpha ^3 m^3$,
  \end{tabular}
\right. \ \ I= \left\{
  \begin{tabular}{cc}
  \ $\frac{128}{3}\mu ^{2}\alpha ^{4}p^{4},$ &
  \ \ \ for {$\alpha \ll \frac{m}{p},$ } \\
  \ $\frac{4}{3} \mu ^2 \alpha ^2 m^2 p^2$, & \ \ \ \ \ \ \ \  { for
  $ \frac{m}{p} \ll \alpha \ll \frac{p}{m},$} \\
\ $4 \mu ^2 \alpha ^4 m^4$, & \ { for
  $ \alpha \gg \frac{p}{m}, $}
  \end{tabular}
\right.
\end{equation}
and in the opposite case, $p\ll m$, we have
\begin{equation}\label{p_ll}
\Gamma = \left\{
  \begin{tabular}{c}
  \ $\frac{64}{3} \mu ^2 \alpha ^3 p^3,$ \\
  \ $\frac{512}{5} \mu ^2 \alpha ^6 p^3$,\\
  \ $4 \mu ^2 \alpha ^3 m^3$,
  \end{tabular}
  \ \ \
I = \left\{
  \begin{tabular}{cc}
  \ $\frac{128}{3} \mu ^2 \alpha ^4 p^4,$ &
  \ \ \ \ \ for {$\alpha \ll 1,$ } \\
  \ $\frac{1024}{3} \mu ^2 \alpha ^8 p^4$, & \ \ \ \ \ \ \ \ \ \ { for
  $ 1 \ll \alpha \ll \frac{m}{p},$} \\
  \ $4 \mu ^2 \alpha ^4 m^4$, & \ \ \ \ { for
  $ \alpha \gg \frac{m}{p}. $}
  \end{tabular}
\right. \right.
\end{equation}
One can see that in the case of a very high matter density the
rate and radiation power are determined by the background matter
density only. Note that the obtained $SL\nu$ rate and radiation
power for $p\gg m$ and $\alpha \gg \frac{m}{p}$  are in agreement
with \cite{Lob_hep_ph_0411342}.
%

From the expressions for the $SL\nu$ rate and total power it is
possible to get an estimate for the average emitted photon energy:
\begin{equation}\label{average_energy}
\left\langle \omega\right\rangle = \frac{I}{\Gamma}.
\end{equation}
In the relativistic case, $p\gg m$,  we get
\begin{equation}\label{overage_omega}
\left\langle \omega\right\rangle \simeq \left\{
  \begin{tabular}{cc}
  \ $2\alpha \frac{p^{2}}{m},$ &
  \ \ \ for {$\alpha \ll \frac{m}{p},$ } \\
  \ $\frac{1}{3} p $, & \ \ \ \ \ \ \ \  { for
  $ \frac{m}{p} \ll \alpha \ll \frac{p}{m},$} \\
\ $\alpha m$, & \ { for
  $ \alpha \gg \frac{p}{m}. $}
  \end{tabular}
\right.
\end{equation}
For the matter parameter $\alpha \gg \frac{m}{p}$ we again
confirm, here, the result obtained in \cite{Lob_hep_ph_0411342}.
In the non-relativistic case, $p\ll m$, we have for the average
emitted photon  energy
\begin{equation}
\left\langle \omega\right\rangle \simeq\left\{
  \begin{tabular}{cc}
  \ $2 \alpha p ,$ &
  \ \ \ for {$\alpha \ll 1,$ } \\
  \ $\frac{10}{3}\alpha^2 p$, & \ \ \ \ \ \ \ \
{ for
  $ 1 \ll \alpha \ll \frac{m}{p},$} \\
\ $\alpha  m $, & \ { for
  $ \alpha \gg \frac{m}{p}. $}
  \end{tabular}
\right.
\end{equation}

\begin{figure}[t]
\begin{minipage}{7.5cm}
\includegraphics[scale=.6]{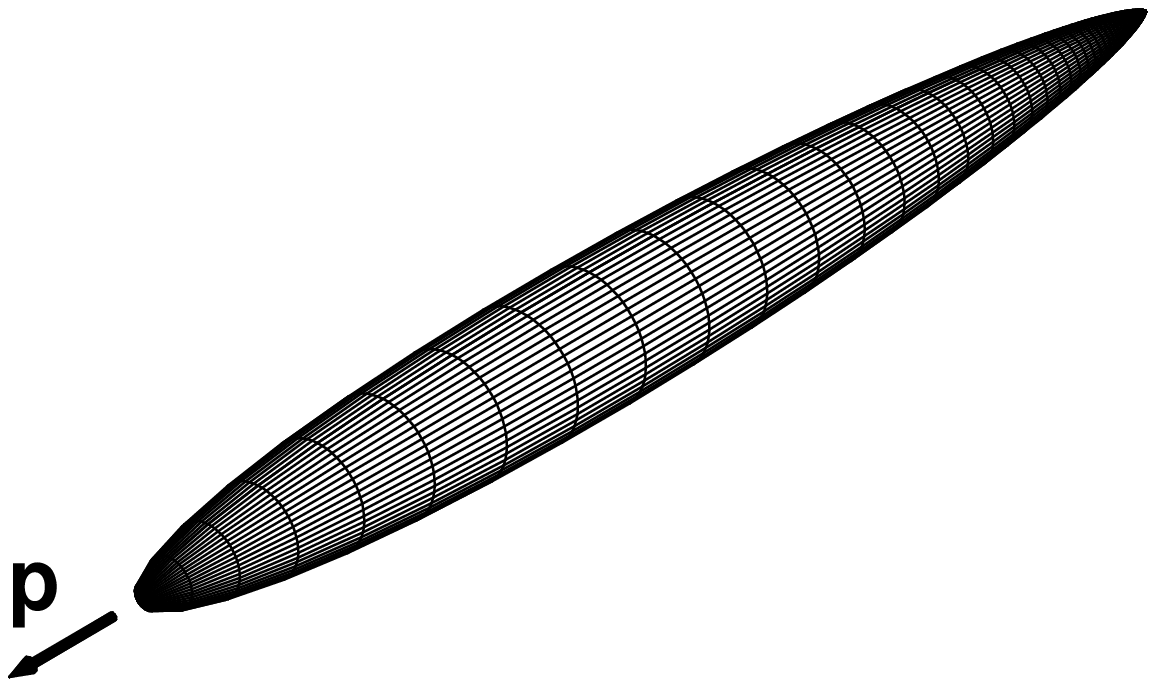}
\caption{The spatial distribution of the $SL\nu$ radiation power
for  $p/m=5, \ \alpha =0.01$.}
\end{minipage}
\hfill
\begin{minipage}{7.5cm}
\includegraphics[scale=.6]{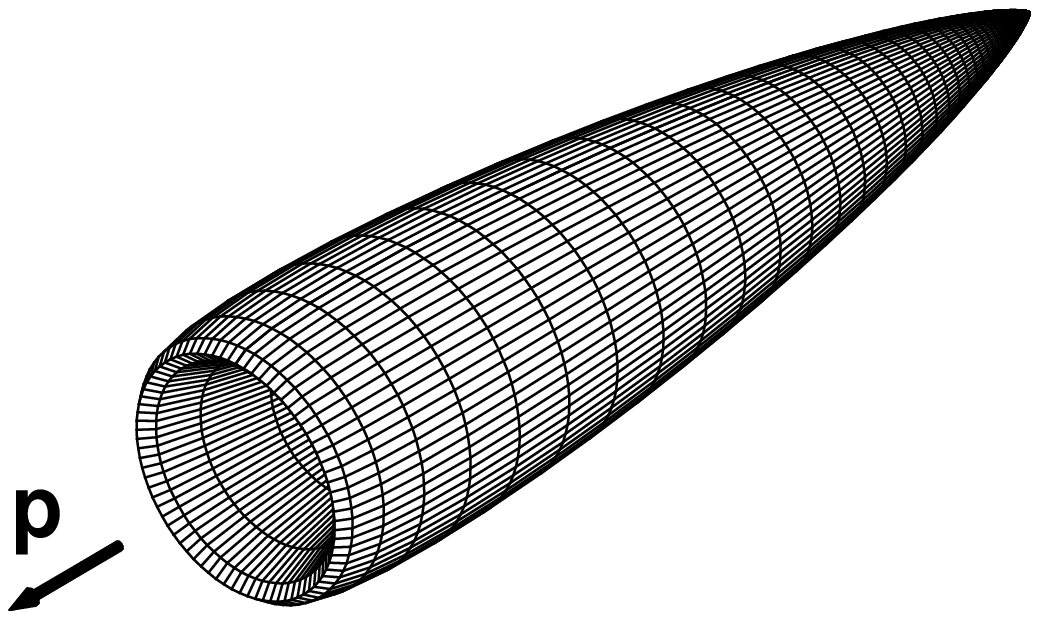}
\caption{The spatial distribution of the $SL\nu$ radiation power
for  $p/m=10^3, \ \alpha =100$. }
\end{minipage}
\end{figure}

We should like to note that for a wide range of neutrino momenta
$p$ and density parameters $\alpha$ the $SL\nu$ power is
collimated along the direction of the neutrino propagation. The
shapes of the radiation power spatial distributions calculated
with use of (\ref{power}) in the case of $p > m$ for low and high
matter density are shown in Figs.2 and 3, respectively. As it
follows from these figures, the shape of the distribution depends
on the density of matter. The shape of the spatial distribution of
the radiation changes from  projector-like to  cap-like with
increase of the matter density. From (\ref{power}) it follows,
that in the case of $p\gg m$ for a wide range of matter densities,
$\alpha\ll \frac{p}{m}$, the direction of the maximum in the
spatial distribution of the radiation power is characterized by
the angle
\begin{equation}\label{angle_max}
\cos \theta_{max} \simeq 1-\frac{2}{3}\alpha\frac{ m}{p}.
\end{equation}
It follows that in a dense matter the $SL\nu$ radiation in the
direction of the initial neutrino motion is strongly suppressed,
whereas there is a luminous ring in the plane per\-pen\-dicular to
the neutrino motion. Note that the rate of the matter-induced
neutrino majoron decay, as it was shown in the second paper of
\cite{BerVysBerSmiPLB87_89}, exhibits a similar angular
distribution.

From analysis of the spatial distribution of the $SL\nu$ radiation
and the emitted photon average energy we predict an interesting
new phenomenon that can appear if a bunch of neutrinos propagates
in a very dense matter. In the case of relativistic neutrinos
$p\gg m$ and dense matter characterized by $\alpha \gg
\frac{p}{m}$ we get that the average value of $\omega \cos \theta$
is negative and equals
\begin{equation}\label {overage_omega_cos}
 \left\langle \omega
\cos \theta \right\rangle=-\frac{1}{3}\alpha m .
\end{equation}
This means that in the considered case a reasonable fraction of
the $SL\nu$ photons are emitted in the direction opposite to the
initial neutrino momentum $p$, as if the neutrinos of the bunch
shake off the spin light photons. It also follows, that in this
case the neutrino momentum $p$ increases as the neutrinos radiate.
To illustrate this phenomena we plot in Fig.4 the $SL\nu$
radiation power spatial distribution for relativistic neutrinos
with $p/m = 10 $ and the density parameter equal to $\alpha =
100$. The two-dimensional cut of the spatial distribution of the
radiation is shown in Fig.5.

\begin{figure}[t]
\begin{minipage}{7.5cm}
\begin{center}
\includegraphics[scale=.6]{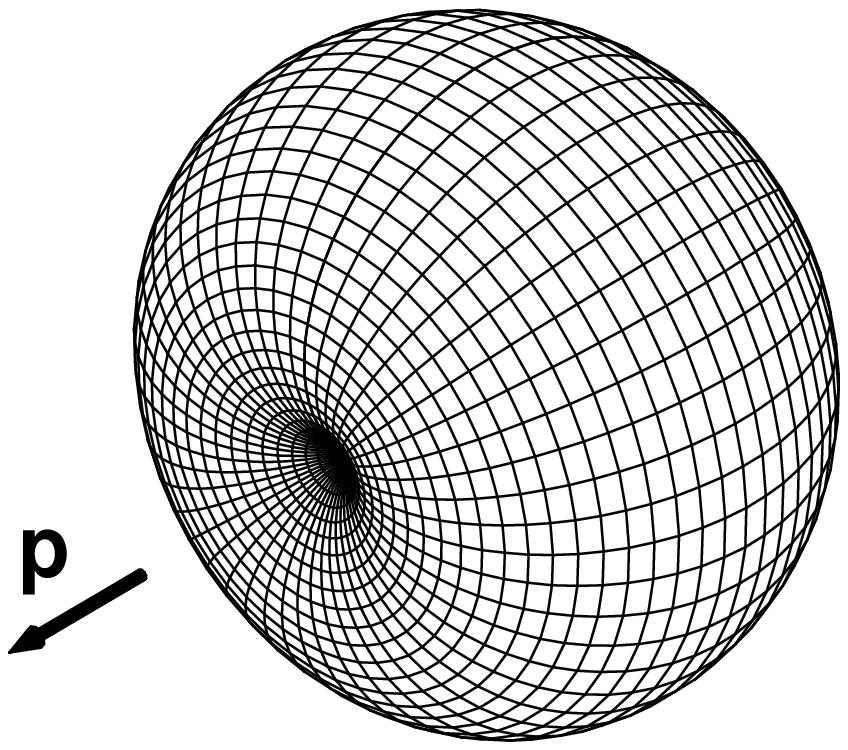}
\caption{The spatial distribution of the $SL\nu$ radiation power
for  $p/m=10, \ \alpha =100$.}
\end{center}
\end{minipage}
\hfill
\begin{minipage}{7.5cm}
\begin{center}
\includegraphics[scale=.4]{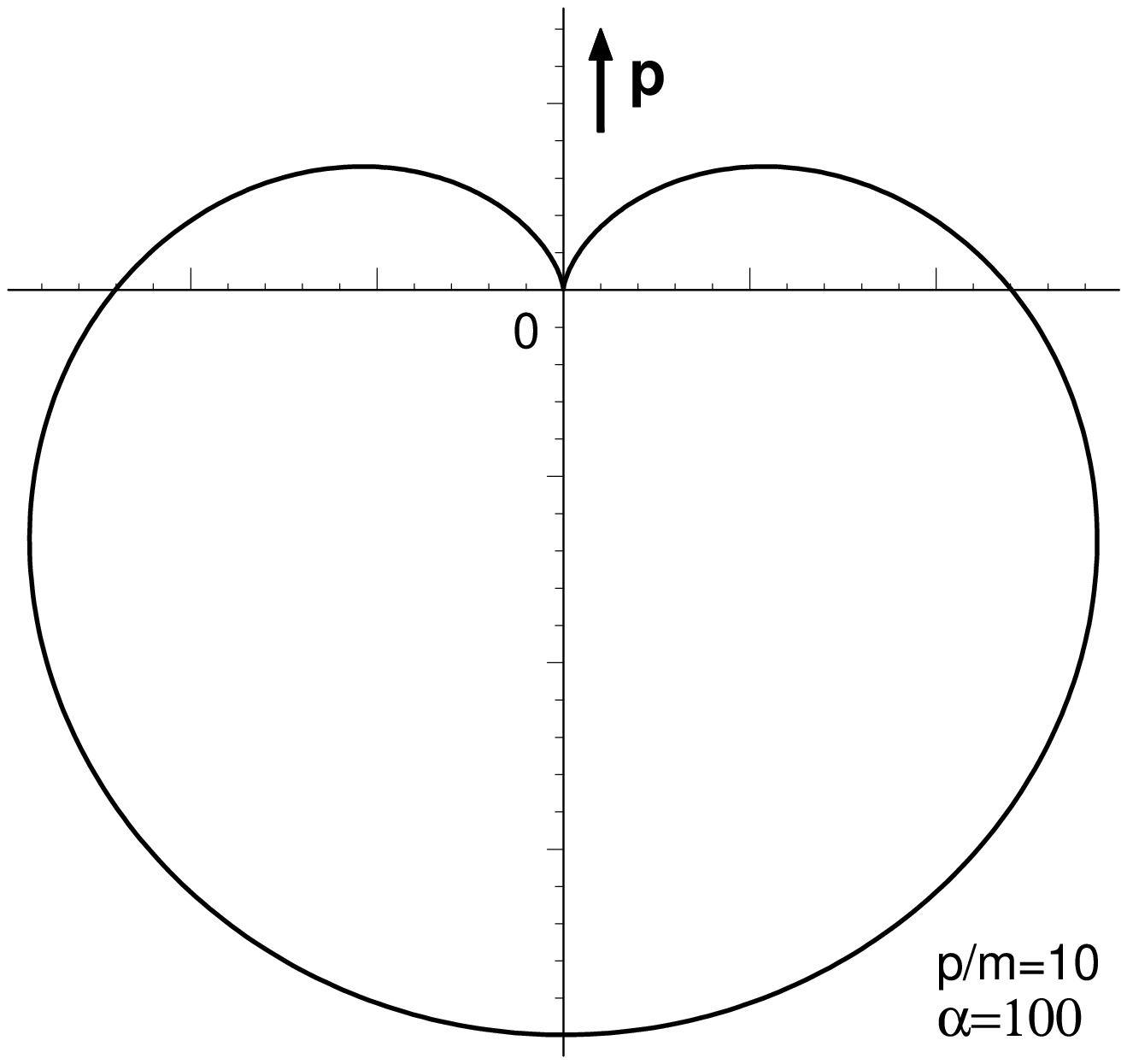}
\caption{The two-dimensional cut of the spatial distribution of
the $SL\nu$ radiation power shown in Fig.4, $p/m=10, \ \alpha
=100$.}
\end{center}
\end{minipage}
\end{figure}

\section{$SL\nu$ polarization properties}

In our previous studies \cite{StuTerPLB05,StuTerQUARKS_04_0410296}
we considered the $SL\nu$ in the low matter density limit,
$\alpha\ll 1$, with account of the photon linear and circular
polarizations. Here, we extend our previous consideration of the
$SL\nu$ polarization properties to the case of an arbitrary matter
density that enables us to treat the emitted photon polarization
in the limit of very high matter density.

We first consider the two different linear photon polarizations
and introduce the two orthogonal vectors
\begin{equation}\label{e_12}
  {\bf e}_1= \frac{[{\bm \varkappa}\times {\bf j}]}
  {\sqrt{1-({\bm \varkappa}{\bf j})^{2}}}, \ \
  {\bf e}_2= \frac{{\bm \varkappa}({\bm \varkappa}{\bf j})-{\bf j}}
  {\sqrt{1-({\bm \varkappa}{\bf j})^{2}}},
\end{equation}
where $\bf j$ is the unit vector pointing in the direction of the
initial neutrino propagation. Decomposing the neutrino transition
amplitude (\ref{amplitude}) in contributions from the photons of
the two linear polarizations determined by the vectors ${\bf e}_1$
and ${\bf e}_2$, we get
\begin{equation}
I^{\left( 1\right) ,\left( 2\right) }=\mu ^2\int_{0}^{\pi
}\frac{\omega ^{4}}{1+\beta ^{\prime }y}\left( \frac{1}{2}S\mp
\Delta S\right) \sin \theta d\theta ,
\end{equation}
where
\begin{equation}
\Delta S=\frac{1}{2}\frac{m^{2}p\sin ^{2}\theta }{\left( E^{\prime
}-\alpha m\right) \left( E-\alpha m\right) p^{\prime }}.
\end{equation}
In the low matter density case, $\alpha \ll 1$, the total
radiation power of the linearly polarized photons is
\begin{equation}
I^{\left( 1\right) ,\left( 2\right) }\simeq \frac{64}{3}\left(
1\mp \frac{1}{2}\right) \mu ^2\alpha ^{4}p^{4}, \label{First}
\end{equation}
in agreement with \cite{StuTerPLB05, StuTerQUARKS_04_0410296}.
Thus, the radiation powers for the two liner polarizations differ
by a factor of three. Contrariwise, in all other cases   the
radiation powers for the two polarizations, ${\bf e}_1$ and ${\bf
e}_2$, are of the same order,
\begin{equation}I^{(1)}\simeq I^{(2)}\simeq
\frac{1}{2}\big(I^{(1)}+I^{(2)}\big).
\end{equation}

It is also possible to decompose the radiation power for the
circularly polarized photons. The two orthogonal vectors
\begin{equation}\label{circ_pol}
  {\bf e}_{l}=\frac{1}{\sqrt 2}({\bf e}_{1}+il{\bf e}_{2})
\end{equation}
describe the two photon circular polarizations ($l=\pm 1$
correspond to the right and left photon circular polarizations,
respectively). For the radiation power of the circular-polarized
photons we obtain
\begin{equation}
I^{\left( l\right) }=\mu ^2\int_{0}^{\pi }\frac{\omega
^{4}}{1+\beta ^{\prime }y}S_{l}\sin \theta d\theta ,
\end{equation}
where
\begin{equation}
S_{l}=\frac{1}{2}\left( 1+l\beta ^{\prime }\right) \left( 1+l\beta
\right) \left( 1-l\cos \theta \right) \left( 1+ly\right).
\end{equation}
In the limit of low matter density, $\alpha \ll 1$, we get for the
power
\begin{equation}
I^{\left( l\right) }\simeq \frac{64}{3}\mu^{2}\alpha
^{4}p^{4}\left( 1-l\frac{p}{2E_{0}}\right).
\end{equation}
In this limiting case the radiation power of the left-polarized
photons exceeds that of the right-polarized photons
\begin{equation}
I^{(-1)}>I^{(+1)}.
\end{equation}
 In particular, this result is also valid for
  non-relativistic neutrinos, $p\ll m$, for a low density with
$\alpha \ll 1$.

It is remarkable that in the most interesting case of  rather
dense matter ($\alpha \gg \frac{m}{p}$ for $p\gg m $ and $\alpha
\gg 1$ for $p\ll m$), the main contribution to the power is
provided by the right-polarized photons, whereas the emission of
the left-polarized photons is suppressed:
\begin{eqnarray}
I^{\left( +1\right) } &\simeq &I, \\
I^{\left( -1\right) } &\simeq &0.
\end{eqnarray}
Thus, we conclude that in a dense matter the $SL\nu$ photons are
emitted with nearly total right-circular polarization. Note that
if the density parameter changes sign, then the emitted photons
will exhibit the left-circular polarization.

\section{Propagation of $SL\nu$ photons in plasma}

Finally, we should like to discuss in some detail restrictions on
the propagation of  $SL\nu$ photons, that are due to the presence
of background electron plasma in the case of $p\gg m$ for the
density parameter  $\frac{m}{p} \ll \alpha \ll \frac{p}{m}$. Only
photons with energy exceeding the plasmon frequency
\begin{equation}\label{pl_freq} \omega _{pl}= \sqrt{\frac{4 \pi
e^2}{m_e} n},
\end{equation}
can propagate in the plasma (here $e^2=\alpha_{QED}$ is the
fine-structure constant and $m_e$ is the mass of the electron).
From (\ref{omega1}) and (\ref{angle_max}) it follows that the
photon energy and the radiation power depend on the direction of
the radiation. We can conclude that the maximal photon energy,
\begin{equation}\label{omega_max}
\omega_{max}=p,
\end{equation}
and the  energy of the photon emitted in the direction of the
maximum radiation power,
\begin{equation}
\omega (\theta_{max})=\frac{3}{4}p,
\end{equation}
are of the same order in the case considered . For  relativistic
neutrinos and rather dense matter the angle $\theta_{max}$, at
which the radiation power (\ref{power}) has its maximum, and the
angle (\ref{omega_max}) corresponding to the maximal photon energy
are both very close to zero (to illustrate this we show in Fig.6
the photon energy and radiation power angular distributions for
the particular case of $m=1 \ eV, p=100 \ MeV$ and $n=10^{32} \
 cm^{-3}$). In addition, as it
follows from (\ref{overage_omega}), the average photon energy
$\left\langle \omega\right\rangle =\frac{1}{3} p$ is also of the
order of $\omega_{max}$ and $\omega (\theta_{max})$. Therefore,
the effective $SL\nu$ photon energy reasonably exceeds the plasmon
frequency  (\ref{pl_freq}) if the following condition is
fulfilled:
\begin{equation}\label{p_in_pl}
p\gg p_{min}=3.5\times 10^{4}\left(
\frac{n}{10^{30}cm^{-3}}\right)^{1/2} eV.
\end{equation}
The $SL\nu$ photon emitted  by a neutrino with momentum $p\gg
p_{min}$ freely propagates through the plasma. For $n \sim 10^{33}
\ cm^{-3}$ we have $p_{min} \sim 1 \ MeV$.

\begin{figure}[t]
\begin{center}
\includegraphics[scale=.45]{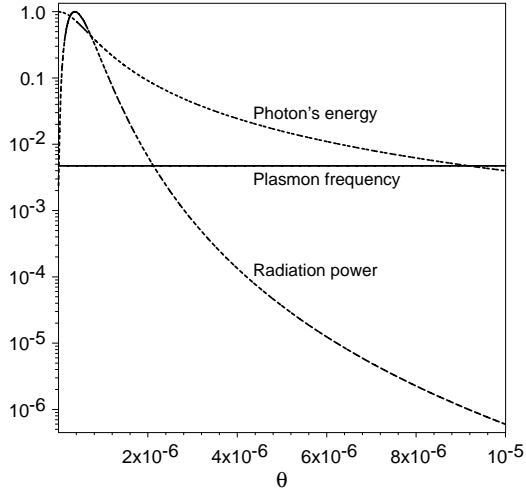}
\caption{The angular dependence  of the emitted photon energy and
radiation power  for the set of parameters: $m=1 \ eV, p=100 \
MeV, n=10^{32} \ cm^{-3}$. The solid line denotes the energy level
that corresponds to the plasmon frequency (\ref{pl_freq}).}
\end{center}
\end{figure}

\section{Summary of $SL\nu$ properties}

 To conclude, we should like to mention that the obtained equation (\ref{Dirac}) is the most general
equation of motion for the neutrino in which the effective
potential accounts for both the charged and neutral-current
interactions with the background matter. Possible effects of the
 motion and polarization of matter can also be  incorporated
\cite{StuTerPLB05,StuTerQUARKS_04_0410296,
GriStuTerCOSMION04_hep_ph0502231}.

The exact solutions obtained  for the modified Dirac equation and
the neutrino energy spectrum form a basis for a rather powerful
method for studying different processes stimulated by neutrinos in
the presence of background matter. For instance, from the neutrino
energy spectrum (\ref{Energy}) and from the matter density
parameters for relativistic electron and muon neutrinos
propagating in matter composed of electrons, protons and neutrons
(see  \cite{StuTerPLB05}) we can get the following expressions for
the two flavour neutrinos:
\begin{equation}
E_{\nu_e, \nu_{\mu}}^{s=-1}\approx E_0 + 2\alpha_{\nu_e,
\nu_{\mu}} m_{\nu_e, \nu_{\mu}},
\end{equation}
where
\begin{equation}\label{alpha-nu-e}
  \alpha_{\nu_e}=\frac{1}{2\sqrt{2}}\frac{G_F}{m}\Big(n_e(1+4\sin^2 \theta
_W)+n_p(1-4\sin^2 \theta _W)-n_n\Big),
\end{equation}
and
\begin{equation}\label{alpha-nu-mu}
  \alpha_{\nu_\mu}=
  \frac{1}{2\sqrt{2}}\frac{G_F}{m}\Big(n_e(4\sin^2 \theta
_W-1)+n_p(1-4\sin^2 \theta _W)-n_n\Big),
\end{equation}
here $n_{e,p,n}$ are the electron, proton, and neutron number
densities, respectively. From the above expressions, the electron
energy shift with respect to the muon energy in matter is
\begin{equation}\Delta E \equiv
E_{\nu_e}^{s=-1}-E_{\nu_{\mu}}^{s=-1} = \sqrt2 G_F n_e.
\end{equation}
Thus, the correct value for the neutrino energy difference
corresponding to the Mikheyev-Smirnov-Wolfenstein effect
\cite{WolPRD78MikSmiYF85} can be recovered.

Now, after the study of $SL\nu$ taking exactly into account matter
effects, performed above, we can summarize the main features of
the phenomena considered  as follows:

1) a neutrino with nonzero mass and magnetic moment emits spin
light when moving in  dense matter,

2) in general, $SL\nu$ in matter is due to the neutrino energy
dependence on the matter density and, in particular, to neutrinos
of the same momentum $p$ but of opposite helicities having
different energies in matter;

3) in the particular case of electron neutrinos moving in matter
composed predominantly of electrons, the matter density parameter
$\alpha$ is positive; here the negative-helicity neutrino (the
left-handed relativistic neutrino $\nu_{L}$) is converted to the
positive-helicity neutrino (the right-handed neutrino $\nu_{R}$),
giving rise to neutrino-spin polarization effect;

4)  the matter density parameter $\alpha$ can, in general, be
negative; therefore the types of initial and final neutrino
states, conversion between which effectively produces the $SL\nu$
radiation, are determined by the matter composition;

5) the obtained expressions for the $SL\nu$ radiation rate and
power , (\ref{Gamma}) and (\ref{power}), exhibit non-trivial
dependence on the density of matter and on the initial neutrino
energy; in particular, as it follows from (\ref{p_gg}) and
(\ref{p_ll}), in the low matter density limit the power is
suppressed by an additional factor of $\frac{m}{p}$ (for $p\gg m$)
or by $\frac{p}{m}$ (for $m\gg p$), in the high density limit,
$\alpha \gg \frac{p}{m}$ ($p\gg m$) or $\alpha\gg \frac{m}{p}$
($m\gg p$), the power acquires the increasing factor $\frac{p}{m}$
(for $p\gg m$) or $\frac{m}{p}$ (for $m\gg p$);

6) for a wide range of matter density parameters the $SL\nu$
radiation is beamed along the neutrino momentum $p$, however the
actual shape of the radiation spatial distribution may vary from
projector-like to cap-like, depending on the  neutrino
momentum-to-mass ratio and the value of $\alpha$;

7) it has been shown that for a certain choice of neutrino
momentum and matter density a reasonable fraction of the emitted
photons move in the direction opposite  to the neutrino momentum
(this interesting phenomenon arises, for instance, in the
particular case of the neutrino parameter $\frac{p}{m} \sim 10$
and $\alpha \sim 100$);

8) in a wide range of matter density parameters $\alpha$ the
$SL\nu$ radiation is characterized by total circular polarization;

9) the emitted photon energy is also essentially dependent on the
neutrino energy and matter density; in particular, the photon
energy  increases from  $\omega \sim 2p$ up to $\omega \sim \alpha
m$ with the density; in the most interesting for astrophysical and
cosmology applications case (when $p \gg m$ and $\frac{m}{p} \ll
\alpha \ll \frac{p}{m}$) the average energy of the emitted photon
 is one third of the neutrino momentum $p$, in the
case of very high density this value equals one half of the
initial neutrino energy in matter.

We argue that the investigated properties of neutrino-spin light
in matter may be important for experimental identification of this
radiation from different astrophysical and cosmological sources.
The fireball model of GRBs (see \cite{ZhaMesIJMP04-PirRMD04} for
recent reviews) is one of the examples. Gamma-rays can be expected
to be produced during collapses or coalescence processes of
neutron stars, owing to the $SL\nu$ mechanism in dense matter
discussed. Another rather favorable situation for effective
$SL\nu$ production   can be realized during a neutron star being
"eaten up" by the black hole at the center of our Galaxy. For
estimation, let us consider a neutron star of mass
$M_{NS}\sim3M_{\bigodot}$ ($M_{\bigodot}=2\cdot10^{33} g$ is the
solar mass). The corresponding effective number density will be
$n\sim8\cdot10^{38} \ cm^{-3}$ and for the matter density
parameter we get $\alpha\sim23$, if the neutrino mass is $m
\sim0.1 \ eV$. For relativistic neutrino energies ($p \gg m$) the
emitted $SL\nu$ photon energy, as it follows from
(\ref{overage_omega}), is $\left<\omega\right>\sim1/3p$, so that
the energy range of this radiation may even extend up to energies
peculiar to the spectrum of gamma-rays. Note that, as it is shown
in Section 4, this radiation is characterized by the total
circular polarization. This fact can be important for experimental
observations.

The authors are thankful to Venyamin Berezinsky, Alexander Dolgov,
Carlo Giunti, Gil Pontecorvo, Victor Semikoz and Alexander
Zakharov for very useful discussions. One of the authors (A.S.)
thank Mario Greco for the invitation to participate in this
Conference and also thanks all the organizers for their kind
hospitality.

\end{document}